\documentclass[lettersize,journal]{IEEEtran}
\usepackage{amsmath,amsfonts}
\usepackage{algorithmic}
\usepackage{algorithm}
\usepackage{array}
\usepackage[caption=false,font=normalsize,labelfont=sf,textfont=sf]{subfig}
\usepackage{textcomp}
\usepackage{stfloats}
\usepackage{url}
\usepackage{verbatim}
\usepackage{graphicx}
\usepackage{cite}
\begin{document}

\title{Thermal Crosstalk Modelling and Compensation Methods for Programmable Photonic Integrated Circuits}%

\author{I. Teofilovic*\thanks{Isidora Teofilovic, Ali Cem, and Francesco Da Ros are with Department of Electrical and Photonics Engineering, Technical University of Denmark, 2800 Lyngby, Denmark (e-mail: isteo@dtu.dk; alice@dtu.dk; fdro@dtu.dk). David Sanchez-Jacome and Daniel Pérez-López are with iPronics Programmable Photonics S.L., Av. Blasco Ibanez 25, 46010 Valencia, Spain (e-mail: david.sanchez@ipronics.com; daniel.perez@ipronics.com )}\thanks{* These authors contributed equally to this work.}, A. Cem*, D. Sanchez-Jacome, D. Pérez-López, and F. Da Ros}%

\markboth{Journal of \LaTeX\ Class Files,~Vol.~14, No.~8, August~2021}%
{Shell \MakeLowercase{\textit{et al.}}: A Sample Article Using IEEEtran.cls for IEEE Journals}

\maketitle

\begin{abstract}
Photonic integrated circuits play an important role in the field of optical computing, promising faster and more energy-efficient operations compared to their digital counterparts. This advantage stems from the inherent suitability of optical signals to carry out matrix multiplication. However, even deterministic phenomena such as thermal crosstalk make precise programming of photonic chips a challenging task. Here, we train and experimentally evaluate three models incorporating varying degrees of physics intuition to predict the effect of thermal crosstalk in different locations of an integrated programmable photonic mesh. We quantify the effect of thermal crosstalk by the resonance wavelength shift in the power spectrum of a microring resonator implemented in the chip, achieving modelling errors $<$0.5 pm. We experimentally validate the models through compensation of the crosstalk-induced wavelength shift. Finally, we evaluate the generalization capabilities of one of the models by employing it to predict and compensate for the effect of thermal crosstalk for parts of the chip it was not trained on, revealing root-mean-square-errors of $<$2.0 pm. 
\end{abstract}


\section{Introduction}\label{chap:intro}
\IEEEPARstart{T}{he} roots of optical computing trace back to the 1960s, with the successful application of incoherent spatial filtering in optical character recognition in 1965 \cite{armitage_character_1965}. The 1980s marked the golden age of optical computing, witnessing numerous applications explored in both academia and industry, including the first attempt in building an optical neural network as early as 1985 in a free-space setting \cite{farhat_optical_1985}. However, progress in digital electronics and the lack of relevant technology, such as integrated photonics, led to a relatively quiet period in optical computing. This changed around a decade ago with the rise of machine learning, and particularly neural networks (NNs), which started to gain significant attention. The simplest form of NNs are  feedforward NNs consisting of artificial neurons organized into interconnected layers. Each layer comprises a linear stage performing matrix-vector multiplication (MVM), and a nonlinear activation function. A vast number of connections among their neurons makes MVM a computationally expensive inference operation. Evolution of NNs has naturally led to a renewed interest in optical computing as optical signals are intrinsically suitable for performing matrix multiplication. Alongside NNs, algorithms for scientific computing, combinatorial optimization, and cryptography rely heavily on MVM, leading to increasing efforts into implementing it by optical means, thus offering improvements in latency, throughput, and energy efficiency \cite{mcmahon_physics_2023}. Optical MVM is usually implemented using photonic integrated circuits (PICs) based on Mach-Zehnder interferometer (MZI) meshes \cite{{shen_deep_2017}, {de_marinis_photonic_2021}}, microring weight banks, \cite{{tait_neuromorphic_2017}, {ashtiani_-chip_2022}, {zhang_silicon_2022}}, semiconductor optical amplifier (SOA)-based architectures \cite{shi_inp_2022}, or photonic crossbar arrays \cite{{youngblood_coherent_2023}, {vlieg_integrated_2022}}. 
 
When designing any computer, several metrics standards need to be met including but not limited to size, robustness, scaling, and accuracy. Many of these are limited by fabrication errors and crosstalk effects in PICs \cite{mcmahon_physics_2023}. A concern for any neural network processor, including analog optical processors, is the potential for the accumulation of errors in deep neural networks, as it has been shown to lead to a 70$\%$ drop in classification accuracy \cite{banerjee_modeling_2021}. Most common approaches to program matrix weights in PICs rely on low-loss thermo-optic phase shifters \cite{de_marinis_photonic_2021}. Deterministic thermal crosstalk between the heaters poses a particular challenge for accurate modelling of PICs, as it was shown in \cite{{perez_multipurpose_2017}, {fang_design_2019}, {bandyopadhyay_hardware_2021}, {cem_data-driven_2023}}. Its influence can lead to a severe degradation in NNs' classification accuracy of up to 85$\%$ even with microheaters being placed few milimeters away from optical components in a photonic chip \cite{biasi_effect_2022}. More generally, accurate programming of PICs for MVMs has been a prominent subject of investigation within the photonics community, encompassing on-line and off-line methodologies, as well as fine-tuning approaches combining both \cite{buckley_photonic_2023}. Off-line methods could ideally offer more general approaches to programming photonic MVM architectures, since their objective is to accurately map the desired matrix onto optical hardware. Consequently, such models are not restricted by specific tasks and do not require additional optimization of heater voltages or on-line calibrations when either a task or a matrix to be implemented changes. This flexibility is particularly valuable in optical MVM proposals that aim to implement large matrices on the same hardware through time-multiplexing submatrices, i.e. continuously reconfiguring the matrix \cite{bruckerhoff-pluckelmann_large_2023}. However, effects such as thermal and electrical crosstalk, and fabrication variations render building an accurate off-line model a complex task.

In this work, we extend the preliminary analysis of \cite{Ali_IPC}, where we introduced 3 off-line crosstalk-predictive models tailored for a programmable photonic mesh comprising 72 MZIs. Crosstalk effect was quantified through a resonance wavelength shift observed in the power spectrum of a microring resonator (MRR) programmed in the mesh. The models were analyzed on a limited experimental dataset and focusing on a single MRR with only few (22) sources of thermal crosstalk positioned in very close proximity to the MRR under test. Here, we conduct a more extensive set of measurements to train the proposed models and extend our investigation to multiple MRRs in the mesh, including contributions of all ($>$ 57) present MZIs. Additionally, we assess the generalization properties of the more physically-informed model by analyzing its cross-predictive performance and validating its efficacy through experimental verification.

The paper is structured as follows: Section \ref{chap:SOTA} gives the overview of the state-of-the-art methods to program and train optical MVM architectures for optical NNs; Section \ref{chap:models} introduces and justifies the specific choice of implementing an MRR in the programmable mesh as a tool to effectively quantify thermal crosstalk; Section \ref{chap:setup&datasets} describes experimental setup and the gathering of the datasets for training the models; Section \ref{chap:modeling_performance} shows the performance of different models when modelling the effect of thermal crosstalk; Section \ref{chap:training_size_analysis} discusses the performance analysis of the models based on data availability;  Section \ref{chap:compensation_performance} shows the generalization performance of one of the models and the experimental compensation results; and Section \ref{chap:conclusion} draws the conclusions. 

\section{State-of-the-art}\label{chap:SOTA}
Different approaches to accurate programming of the PICs for optical MVM  (for NNs) have been introduced so far, including both on-line (in situ) and off-line (in silico) methods. In this context, it is important to distinguish between two key terms: programming and training. Programming involves configuring a circuit to execute a specific (linear) operation. For instance, programming an MZI mesh entails setting driving voltages to ensure the mesh replicates a desired matrix. Training refers to the iterative process of optimizing weight values in an NN to achieve optimal performance, such as maximum classification accuracy or minimum regression errors. The terms off-line and on-line apply to both training and programming. 

\subsection{On-line methods}
Online methods encompass both online training of the optical NNs and online programming PICs to implement a desired linear transformation.

The on-line training methods can be broadly categorized into those employing backpropagation (BP) on the chip, more biologically-plausible BP-free methods, and methods developed specifically for deep physical NNs. The method for calculating gradients necessary for implementing BP using the adjoint variable method, i.e. measuring the interference intensity of the propagating adjoint field and its time-reversed copy was proposed in \cite{hughes_training_2018}. Challenges in enhancing the efficiency of this algorithm restrict its suitability to handling at most 100 MZIs \cite{gu_light_2022}. The hybrid on-line/off-line physics-aware training method presented in \cite{wright_deep_2022} relies on performing a forward pass through a physical system and implementing BP to train controllable physical system digitally by approximating the gradients. By applying this approach, it is not necessary to have complete knowledge of the exact gradients (i.e., the exact physical learning system) in order to train the NN. On-line optical BP using infrared cameras to monitor optical power propagating through any phase shifter in the circuit was proposed in \cite{pai_experimentally_2023}. Such methods usually require additional monitoring hardware and do not scale well with the chip size.

BP being recognized as a non-biologically plausible algorithm for training neural networks \cite{bengio_towards_2015} led to increasing efforts into developing more biologically-plausible training algorithms, such as direct feedback alignment (DFA) \cite{Nkland2016DirectFA} and forward-forward algorithm (FFA) \cite{hinton2022forwardforward}. An auxiliary circuit for training an optical NN with DFA was suggested in \cite{filipovich_silicon_2022}. The method is based on MVM using arrays of MRRs to calculate gradients for each of the NN layers on-line. Examples of applying FFA to optical NNs can be found in in \cite{{momeni_backpropagation-free_2023}, {oguz_forwardforward_2023}}. Besides implementing BP and BP-free algorithms developed for digital NNs on physical hardware, there have been proposals of algorithms for training deep physical NNs. A model-free perturbative method for training arbitrary physical (both digital and analog) NNs was proposed in \cite{buckley_general_2022}. Not relying on backpropagating the error makes these training methods intrinsically more hardware-friendly for optical NNs, yet they are still device- and task-specific, and often rely on complex digital iterative optimization. 

The on-line programming is based on fine-tuning and online-chip calibration to implement a desired linear transformation. A deterministic approach proposed in \cite{bandyopadhyay_hardware_2021} utilizes locally correcting hardware errors with individual optical gates to correct circuit errors. Authors in \cite{zhang_silicon_2022} reported a 2-bit increase in MVM bit precision using dithering signal to stabilize programmed MRR arrays for desired weights. These methods commonly require additional monitoring, and re-optimizing whenever the new linear transformation (i.e. matrix) needs to be implemented.

\subsection{Off-line methods}
Off-line training methods entail training an NN model on a digital computer and transferring the trained network weights to a physical hardware, while off-line programming methods assume building a model accurately describing the PIC behaviour.

The majority of off-line programming models can be broadly split into physics-based and data-driven models. A two step software-based training scheme to configure phase shifts in an MZI-based feedforward optical NN, comprising a stochastic gradient descent algorithm followed by a genetic algorithm introducing four types of practical imprecisions in MZIs (phase shift error, insertion loss, drift of the coupling coefficient, and photodetection noise) was proposed in \cite{shao_generalized_2022}. The effect of thermal crosstalk was neglected, even though it’s been shown that it can cause a degradation of up to 85\% in the NN classification accuracy \cite{biasi_effect_2022}. The authors in \cite{biasi_effect_2022} modeled the heat flux the microheaters generate and modified accordingly the non-linear equations used to describe the MRR system used to carry out both linear and nonlinear operations. A hardware-aware training model where a photonically implemented non-linear activation function, and noise- and bandwidth-induced limitations are a priori included in the training process was introduced in \cite{moralis-pegios_neuromorphic_2022}. A framework for modelling and optimizing photonic NNs was suggested in \cite{shah_analogvnn_2023}. Implementing component imprecisions into models was introduced in \cite{{fang_design_2019}} and \cite{banerjee_modeling_2021}, too. In \cite{Sarantoglou_2022}, a Bayesian training strategy is presented to reduce the influence of minimal thermal crosstalk on performance; however, residual effects are still present. A data-driven approach to modelling an MZI-mesh using NNs was presented in \cite{cem_data-driven_2023}. However, the scalability of the models remains an issue.
Besides these, a training mechanism for optical NNs based on MZI meshes has been proposed in \cite{lu_efficient_2023}. Implementing an arbitrary matrix using MZI meshes requires performing singular value decomposition (SVD) to decompose the matrix to its unitary components and implement the unitary matrices using MZI-meshes. That motivated developing a training algorithm to assure that weight matrices are unitary matrices in each iteration. Subsequently, the number of MZIs needed to implement trained network weights onto physical hardware could be lower, which could result in the less pronounced effect of thermal crosstalk, as there would be less phase-shifters present in the chip. However, it does not fully alleviate the crosstalk issue which would still be present when scaling to large MVMs' sizes.

\section{Approaches for modelling the effect of thermal crosstalk}\label{chap:models}

\begin{figure*}[!t]
    \centering
    \includegraphics[width=\textwidth]{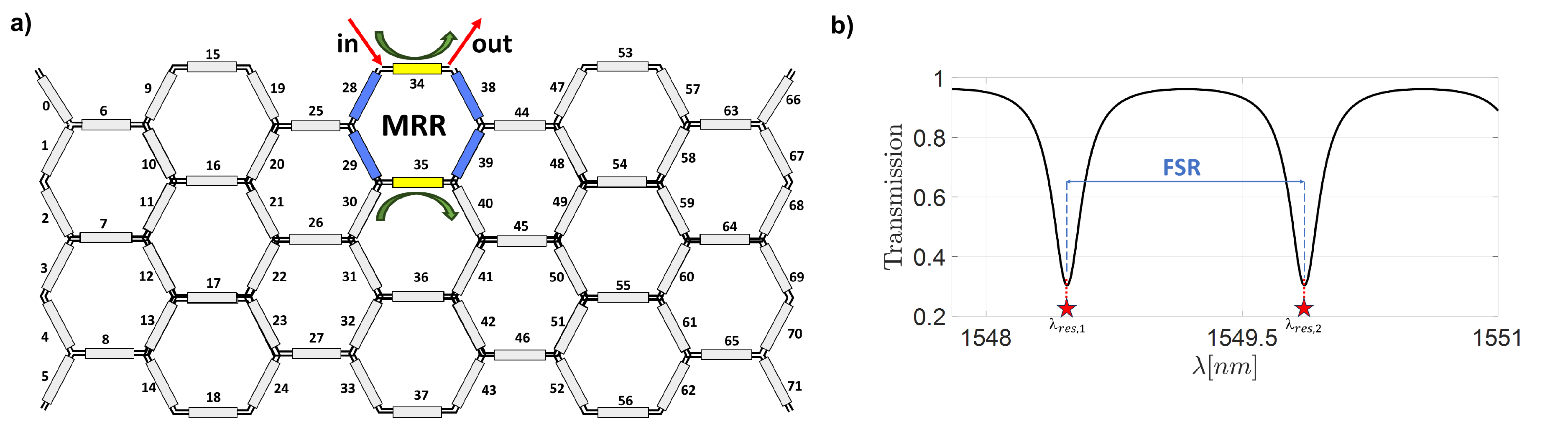}
    \caption{a) Schematic of the 72-PUC waveguide mesh with the MRR filter. Blue-colored PUCs  denote PUCs in bar-state. Yellow-colored PUCs denote PUCs used as tunable couplers. The green arrows indicate the simulation of the coupling waveguides. Red arrows indicate input (in) and output (out) ports. b) Spectral response of an MRR filter. Resonance wavelengths are marked with red stars. FSR denotes the period of the MRR's spectral response.}
    \label{fig:mesh_MRR_spectrum}
\end{figure*}

\subsection{Quantifying thermal crosstalk}
The chip in this investigation is a silicon-based integrated programmable photonic processor (Smartlight, iPRONICS). It comprises an hexagonal mesh of 72 individually controlled programmable unit cells (PUCs), as depicted in Figure \ref{fig:mesh_MRR_spectrum} a). Each PUC is an MZI with two thermo-optic phase shifters, one on each arm. Two fixed couplers in each MZI have close to a 50:50 splitting ratio, enabling the realization of all power coupling ratios from $0$ (bar-state) to $100\%$ (cross-state). By appropriately programming the PUCs, various functionalities can be achieved, including programming different linear transformations, i.e. matrices \cite{perez-lopez_general-purpose_2024}. As discussed in Section \ref{chap:SOTA}, thermal crosstalk has a significant impact on optical MVMs, especially when considering large and footprint-efficient meshes.

The mesh consisting of hexagonal blocks and the physics of the MRRs make it convenient to program the MRR in the mesh and quantify the crosstalk effect by the induced resonance wavelength shift ($\Delta \lambda_{res}$) in the MRR's spectral response. Optical path of an MRR is the product of the effective refractive index of the waveguide ($n_{eff}$) and the MRR round-trip length ($L$). When the optical path length of the MRR is equal to an integer multiple of the wavelength, i.e. $n_{eff}L=m\lambda$ ($m \in \mathbb{Z}$), the MRR is in resonance. For resonance wavelengths, the transmission through the bus waveguide is minimized. The response of the MRR filter is therefore periodic in wavelength with the period being called free spectral range (FSR) and being equal to $FSR=\frac{\lambda^2}{n_gL}$, as shown in Figure \ref{fig:mesh_MRR_spectrum} b), with $n_g$ being a group refractive index \cite{bogaerts_silicon_2012}. The most common way of adjusting the resonance wavelength of an MRR relies on thermal tuning \cite{bogaerts_silicon_2012}. Total phase shift accumulated in one round-trip of the MRR ($\Phi=\frac{2 \pi n_{eff}}{\lambda}L$) is adjusted using thermal heaters. This leads to a change in refractive index of the waveguide ($n_{eff}$), thus in its resonance wavelength. Simultaneously, the increase in temperature around the MRR (due to additional thermal actuators present in the chip) results in additional optical signal delay - the effect of thermal crosstalk. Change in the resonance wavelength of an MRR can be analytically estimated by \eqref{eq:lambda_res} \cite{jacques_optimization_2019}, \cite{lydiate_modelling_2017}, \cite{padmaraju_resolving_2014}:

\begin{equation}
    \Delta \lambda_{res} = \frac{\lambda_{res}}{n_{g0}}(\frac{\delta n_{eff}}{\delta T}dT+n_{g0}\frac{1}{L}\frac{\delta L}{\delta T}dT)\quad
    \label{eq:lambda_res}
\end{equation}

In the equation above, $\lambda_{res}$ is the resonance wavelength, $n_{g0}$ is the group index of the waveguide at that wavelength, and $T$ is the temperature.

The add-drop MRR configuration can be achieved by programming 6 PUCs in the mesh to form a hexagon MRR-like structure, as shown in Figure \ref{fig:mesh_MRR_spectrum} a). Each PUC has a basic unit length of 811.41 $\mu$m, corresponding to the FSR of 118.4 pm. The PUCs programmed as tunable couplers in Figure \ref{fig:mesh_MRR_spectrum} a) perform the coupling by implementing a floating couplig factor. Therefore, the effect of thermal crosstalk was observed by programming MRR filters in the mesh and quantifying $\Delta \lambda_{res}$ in their output power spectra while tuning the phase shifts driven to the PUCs outside the MRRs ("interfering" PUCs).
 
Although not strictly an MRR, assessing the impact of thermal crosstalk through $\Delta \lambda_{res}$ in the hexagonal MRR structure can offer valuable insights into thermal crosstalk effects on the chip. The method could potentially provide a general tool for quantifying and mitigating thermal crosstalk effects within the tested chip, irrespective of the specific task, architecture, or functionality implemented using chip.

\subsection{Thermal crosstalk-predictive models}
As previously shown in  \eqref{eq:lambda_res}, $\Delta \lambda_{res}$ is proportional to the temperature change around the MRR, thus to the dissipated thermal energy from the interfering PUCs once the thermal equilibrium has been established. Although linear dependence of $\Delta \lambda_{res}$ on the phase shifts driven to the interfering PUCs $\Phi_{i}$ could be expected, fabrication tolerances and the complex chip geometry make building a purely physics-based model a challenging task. Therefore, we propose 3 models that relate phases driven to the $N$ interfering PUCs to $\Delta \lambda_{res}$. The models are built incorporating different levels of physical intuition.

\subsubsection{Total phase model (TPM)}
From a physics perspective, it is anticipated that the contribution of each PUC depends linearly on the phase-shift applied to it. Consequently, the initial model introduced is a TPM, which presumes a linear relationship with the phase-shifts of the PUCs while assuming they are all equidistant. The model is given by \eqref{eq:TPM}:

\begin{equation}
    \Delta \lambda_{res} = s\sum_{i=1}^{N} \Phi_i.
    \label{eq:TPM}
\end{equation}

The model comprises a single fitting parameter - the scaling factor $s$. This is the simplest of all the models, and its performance mainly serves as the baseline for more complex ones.

\subsubsection{Thermal decay model (ThDM)}
ThDM assumes a linear dependence of $\Delta \lambda_{res}$ on the phase shifts $\Phi_{i}$, but takes into consideration physical proximity of each PUC to the MRR center. Inspired by thermal diffusion assuming a uniform medium, the model is given by \eqref{eq:ThDM}.

\begin{equation}
    \Delta \lambda_{res} = \sum_{i=1}^{N}(p_1e^{-p_2 d_i}+p_3 d_i+p_4) \Phi_i.
    \label{eq:ThDM}
\end{equation}

Since the contribution of each of the interfering PUCs is expected to follow an exponential decay, fitting parameter $p_2$ is expected to be positive. Considering that ThDM takes into account physical distance ($d_i$) from the PUCs to the MRR under test, and comprises 4 fitting parameters ($p_1, ..., p_4$ in \eqref{eq:ThDM}), it is expected to perform better than the TPM.

\subsubsection{Linear regressor (LR)}
Keeping the assumption of linear dependence on each PUC phase shift independently, but removing any assumption on its dependence on its distance to the MRR center, leads from the ThDM to a LR model, where each of the contributions is learned solely from the experimental data. Therefore, the model corresponds to \eqref{eq:LR}:

\begin{equation}
    \Delta \lambda_{res} = \sum_{i=1}^{N} a_i \Phi_i,
    \label{eq:LR}
\end{equation}

with $a_i,$ $i=1, ..., N$ being the LR model weights. 

\section{Experimental setup and dataset generation}\label{chap:setup&datasets}

\begin{figure}
    \centering
    \includegraphics[width=\columnwidth]{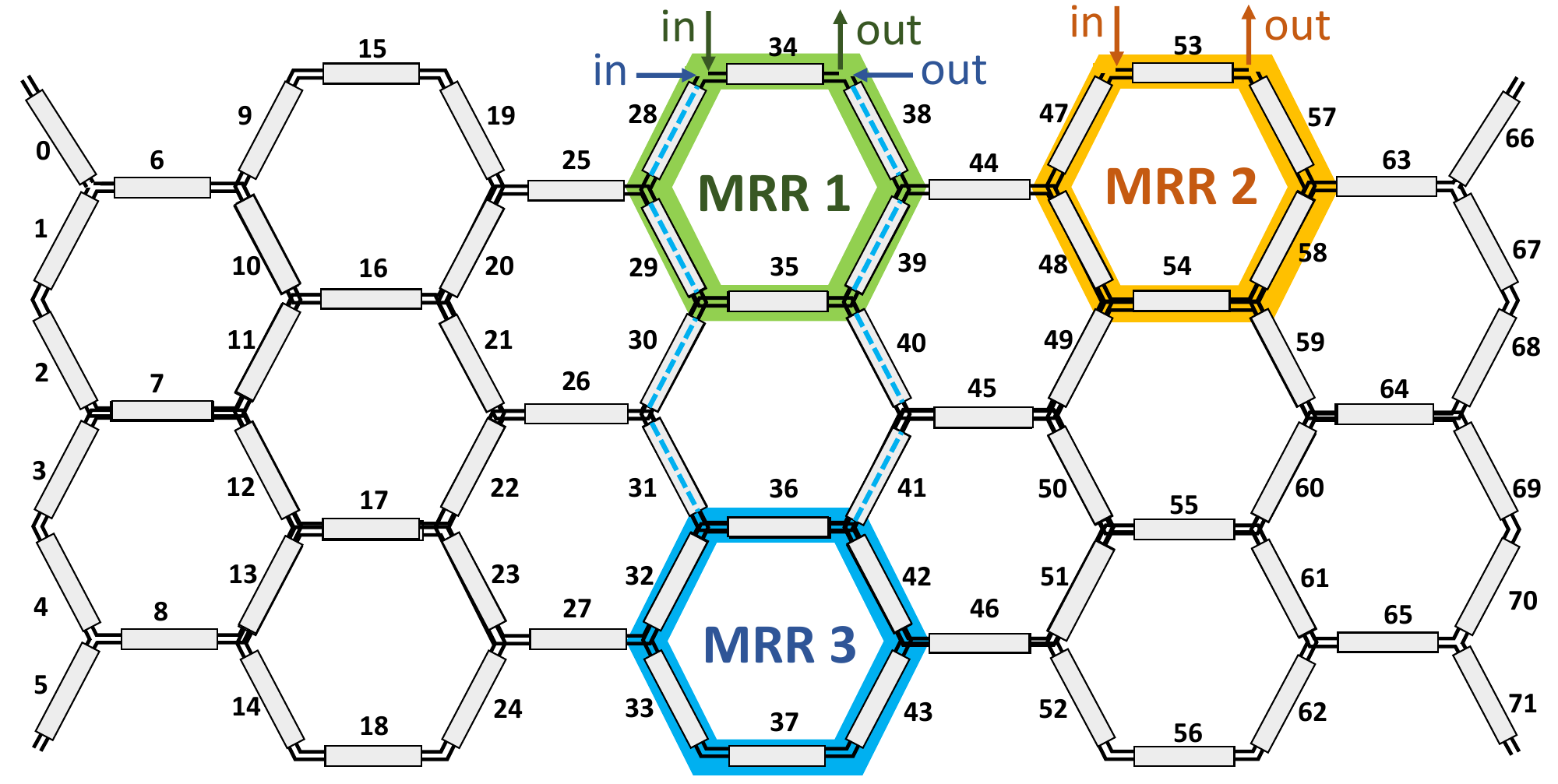}
    \caption{Hexagonal mesh with 3 MRR filters used for thermal crosstalk modelling. MRR 1 (green), MRR 2 (orange), and MRR 3 (blue) are indicated using different colors. Coupling ratios of the PUCs coupling to the input and output ports (34, 36, and 53 for MRRs 1-3, respectively) were set to 0.9. Maximum ERs were achieved by setting coupling ratios of the PUCs 35, 37 and 54 (in MRR 1, MRR 2, and MRR 3, respectively) to 0.77. Achieved ER values were 25, 20, and 30 dB for MRRs 1-3, respectively. The rest of the PUCs forming the MRRs were set to the bar state. Programming MRR 3 required fixing PUCs accross the blue dashed line to guide the light to the input and the output port. Therefore, PUCs 31 and 41 were set to the bar-state, and PUCs 28, 29, 30, 38, 39, 40 were set to the cross-state.}
    \label{fig:HexMesh_MRRs}
\end{figure}

MRRs under investigation are depicted in Figure \ref{fig:HexMesh_MRRs}. All of them were programmed to form an add-drop filter. 

The TPM and the LR models take as inputs interfering PUCs' phase shifts, whereas the ThDM incorporates both phase shifts and distances from the PUCs to the MRR center. The distances were computed using software provided with the chip, which offers information on the distances between any two PUCs in the mesh. The approximate distance between a given PUC and the MRR center was calculated as a mean of distances from the PUC to the six PUCs forming the MRR loop. This distance information was then utilized when training the ThDM.

Tuning the phase shift of a single interfering PUC does not result in a significant resonance wavelength shift by itself. Therefore, after programming the MRR, phase shifts of all the interfering PUCs were simultaneously tuned  to observe the effect of thermal crosstalk. Setting both phase shifters in a given PUC to the same value ensured that the coupling ratio remained constant, assuming that both of the phase shifters are equally affected by crosstalk due to their vicinity. The phase shifters were simultaneously tuned to different random values derived from beta distribution, scaled by $2\pi$ to fit the range [$0, 2\pi$]. The probability density function (PDF) of beta distribution for a random variable $X$ is given in \ref{PDF_beta}:

\begin{equation}
\label{PDF_beta}
f(x; \alpha \beta)=c(\alpha,\beta)x^{\alpha-1}(1-x)^{\beta-1}
\end{equation}

$\alpha$ and $\beta$ are the shape parameters, $x$ is a realization of the random variable $X$ and $c$ is a constant that depends on the shape parameters to ensure that the total probability is within [0, 1]. The distribution can also be parameterized in terms of its mean $\mu$ and the standard deviation $\sigma$ as shown in \eqref{eq:v} to \eqref{eq:beta}:

\begin{equation}
    v=\frac{\mu(1-\mu)}{\sigma^2}-1
    \label{eq:v}
\end{equation}

\begin{equation}
    \alpha=\mu\sigma
    \label{alpha}
\end{equation}

\begin{equation}
    \label{eq:beta}
    \beta=(1-\mu)v
\end{equation}

\begin{figure}
    \centering
    \includegraphics[width=\columnwidth]{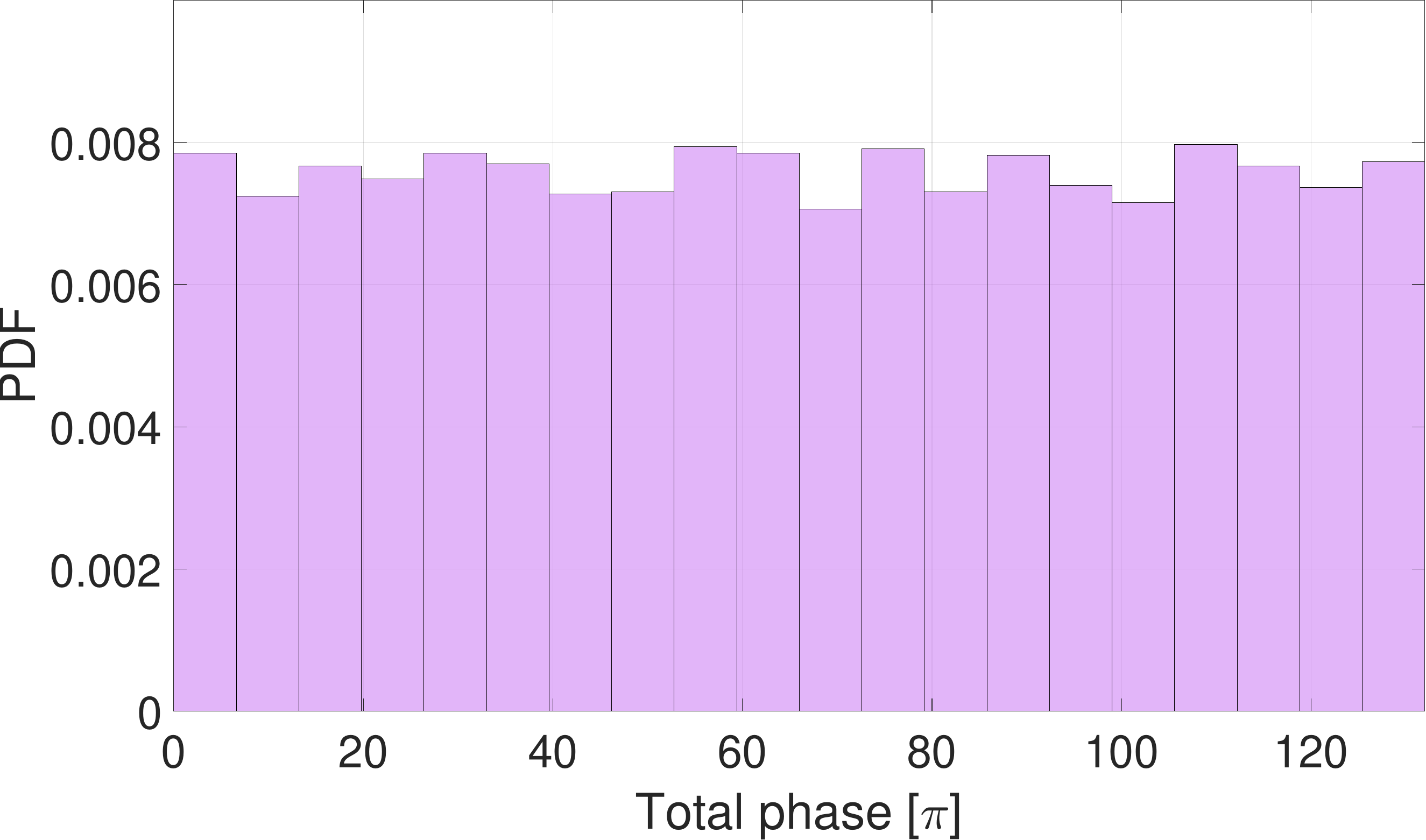}
    \caption{Histogram of the distribution of the total driven phase on the PUCs}
    \label{beta_phases}
\end{figure}

The intermediate parameter $v$ is the sample size of the distribution and $\sigma^2$ was set to 0.05. To assure an even distribution for the total phase, the mean of the beta distribution was not chosen to be a fixed parameter. Instead, the range for the total driven phase $[0, 132\pi]$ was divided into 20 portions. The mean $\mu$ was set to the center of each portion for all distributions when generating the driven phases for the PUCs. An equal number of phases were generated for each portion, resulting in an even coverage of the total range as shown in Figure \ref{beta_phases}.

\begin{figure*}
    \centering
    \includegraphics[width=\textwidth]{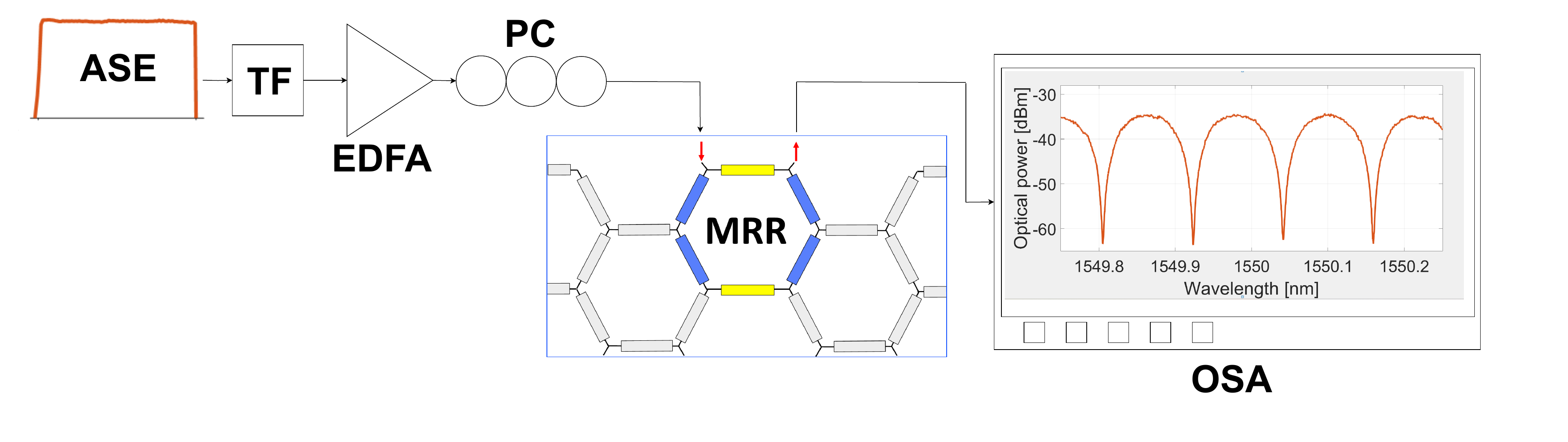}
    \caption{Experimental setup for thermal crosstalk observation. ASE: amplified spontaneous emission, TF: tunable filter, EDFA: Erbium-doped fiber amplifier, PC: polarization controller, OSA: optical spectrum analizer.}
    \label{fig:exp_setup}
\end{figure*}

Having defined the phases, the next step was to perform the measurements and record the output power spectra. The output of a polarized amplified spontaneous emission (ASE) source was filtered to a 1-nm bandwidth centered at 1550 nm with a tunable filter, and fed to an Erbium-doped fiber amplifier (EDFA). Polarization of the EDFA output was optimized using a polarization controller (PC) and then fed to the processor. The power spectra of the MRRs were monitored by optical spectrum analyzer (OSA) within a range of [1549.75, 1550.25] nm with a nominal wavelength resolution of 1.6 pm. 5000 measurements per MRR were performed, with the thermal equilibrium being established before each measurement. The experimental setup is depicted in Figure \ref{fig:exp_setup}.

The measured power spectra were upsampled using spline interpolation to have a step-size of 0.01 pm in wavelength. Each power spectrum was then compared with the “reference spectrum”, which was the MRR spectral response when no additional phase shifts were driven to any of the interfering PUCs. The reference spectrum was measured before each measurement. $\Delta \lambda_{res}$ was then found as a mean of $\Delta\lambda_1$ and $\Delta\lambda_2$. $\Delta\lambda_1$ is a wavelength shift between the shifted spectrum and the reference spectrum measured before it. $\Delta\lambda_2$ is a wavelength shift between the shifted spectrum and a reference spectrum measured after it. $\Delta\lambda_1$ and $\Delta\lambda_2$ were determined as the resonance wavelength shifts that yielded the highest correlation between the shifted spectrum and the corresponding reference spectrum. This approach averages out possible short-term temperature drifts. Only wavelength shifts within 1 FSR were considered when maximizing the correlation. As such, datasets consisting of vectors of phases driven to the interfering PUCs and the resulting $\Delta \lambda_{res}$ were obtained for each of the MRRs. The datasets underwent an $80\% -20\% $ split for training and testing the models.

\section{Modelling performance}\label{chap:modeling_performance}
Training and testing root-mean-square errors (RMSE) for all the models for all the MRRs are given in Table \ref{tab:table1} and Table \ref{tab:table2}, respectively.

\begin{table}[!t]
\caption{Training RMSE $[pm]$\label{tab:table1}}
\centering
\begin{tabular}{|c||c||c||c|}
\hline
  & MRR 1 & MRR 2 & MRR 3\\
\hline
TPM & 0.52 & 0.54 & 0.34\\
\hline
ThDM & 0.41 & 0.34 & 0.25\\
\hline
LR & 0.33 & 0.26 & 0.20\\
\hline
\end{tabular}
\end{table}

\begin{table}[!t]
\caption{Testing RMSE $[pm]$\label{tab:table2}}
\centering
\begin{tabular}{|c||c||c||c|}
\hline
  & MRR 1 & MRR 2 & MRR 3\\
\hline
TPM & 0.52 & 0.53 & 0.34\\
\hline
ThDM & 0.42 & 0.34 & 0.27\\
\hline
LR & 0.34 & 0.27 & 0.21\\
\hline
\end{tabular}
\end{table}

\subsection{TPM}
\begin{figure}
    \centering
    \includegraphics[width=\columnwidth]{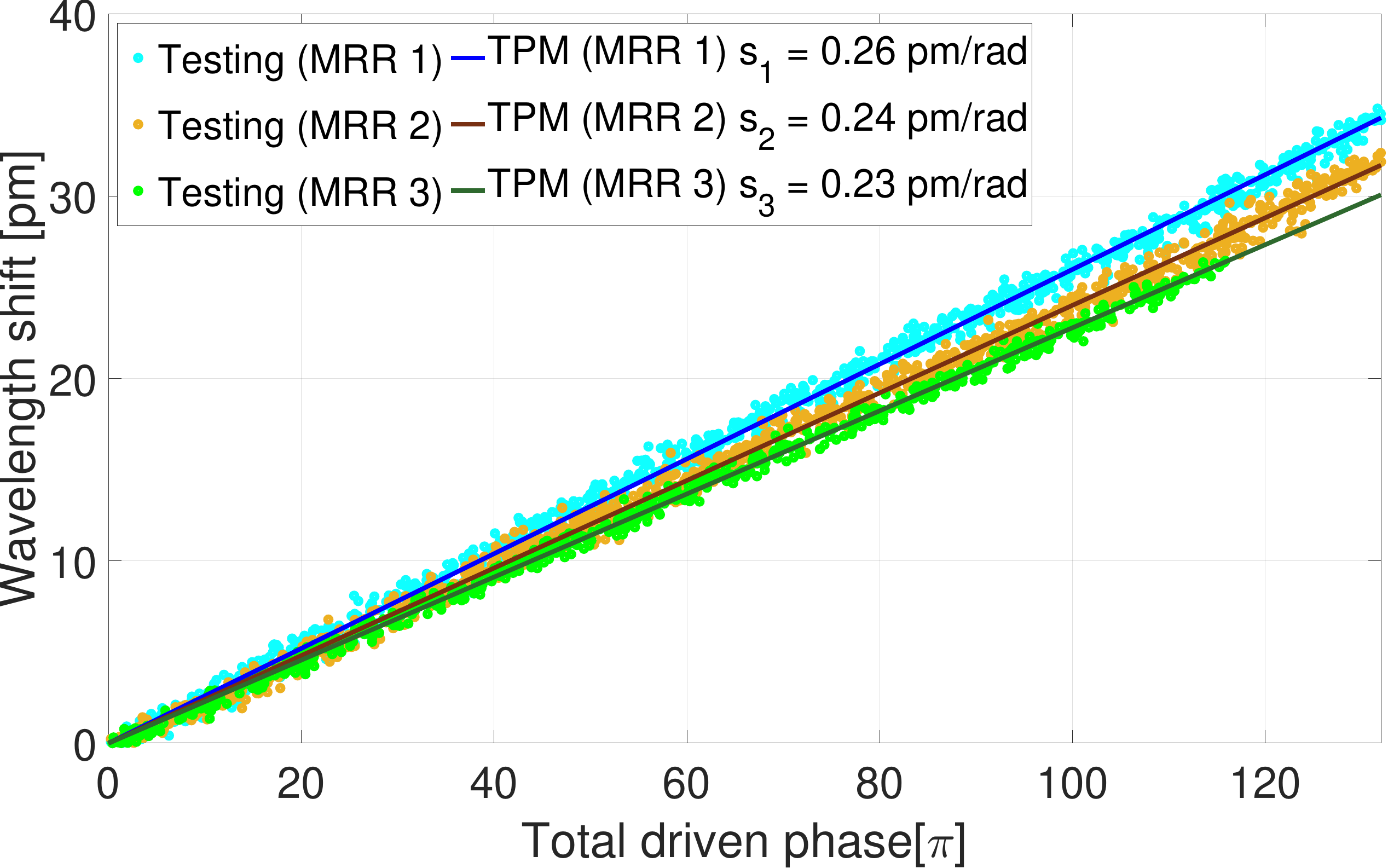}
    \caption{Total Phase Model fitted for for MRR 1, MRR 2, and MRR 3}
    \label{fig:TPM}
\end{figure}

The training RMSEs are found to be 0.52 pm, 0.54 pm, and 0.34 pm; and the testing RMSEs are found to be 0.52 pm, 0.53 pm, and 0.34 pm for MRR 1, MRR 2 and MRR 3, respectively. The optimal value for the scaling factor $s$ are $0.26\frac{pm}{\pi}$, $0.24\frac{pm}{\pi}$, and $0.23\frac{pm}{\pi}$ for MRR 1, MRR 2, and MRR 3, respectively. It is remarkable how well this simple model performs at predicting $\Delta\lambda$ even though it treats all PUCs equally. Nonetheless, by merely summing the phases, the input space is reduced to the dimension of 1, thereby constraining the maximum achievable performance to that of the best-fitting lines depicted in Figure \ref{fig:TPM}.

\subsection{ThDM}
\begin{figure}
    \centering
    \includegraphics[width=\columnwidth]{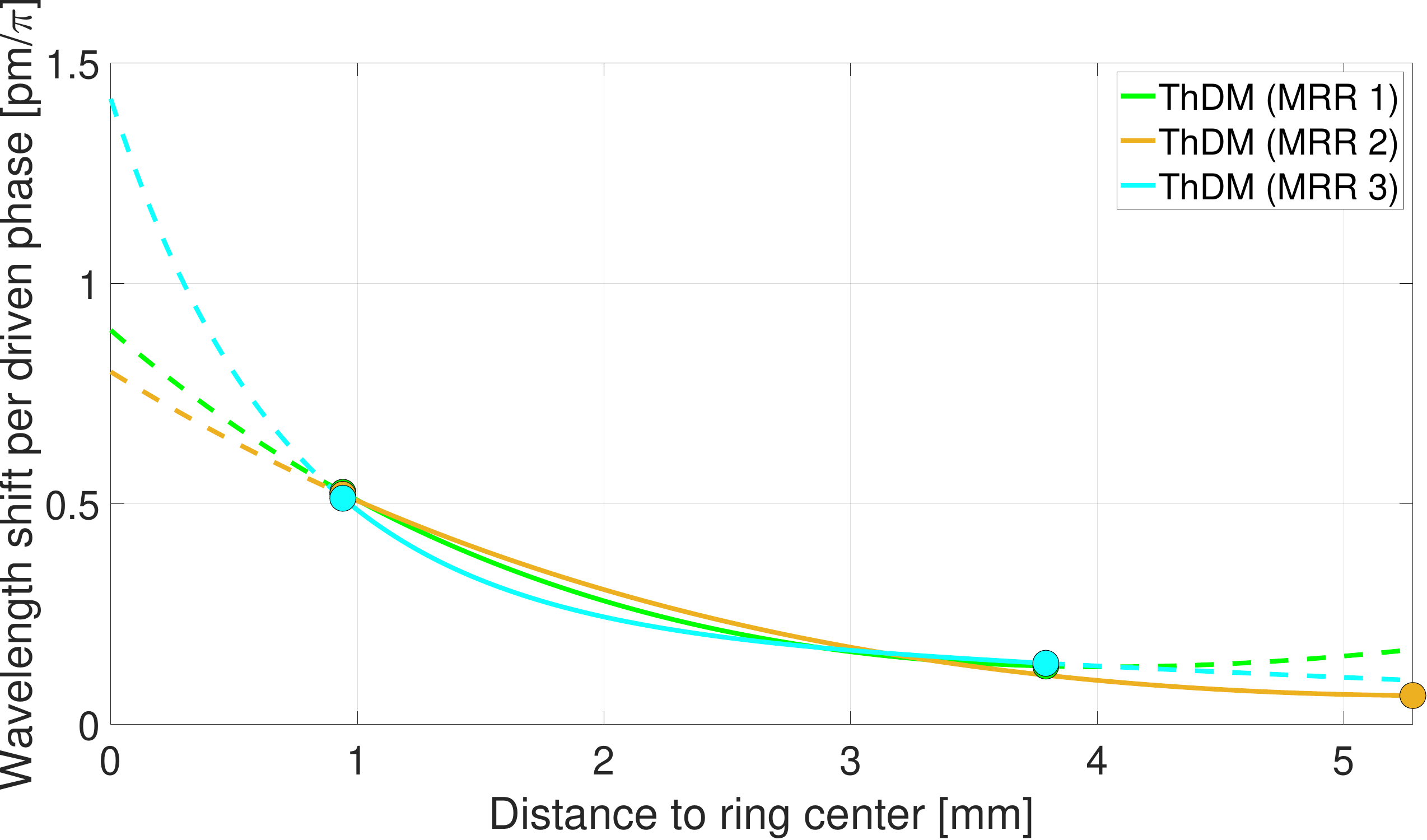}
    \caption{Thermal Decay Model fitted for all 3 MRRs. The dashed portions of the curves show the extrapolation for PUC distances not present in the chip. Circle markers denote minimum and maximum distance present in each MRR dataset.}
    \label{fig:ThDM}
\end{figure}

Taking distances into account and having 4 degrees of freedom compared to a single one in case of the TPM leads to the training and testing RMSEs dropping to 0.41 pm, 0.34 pm, 0.22 pm; and 0.42 pm, 0.34 pm, and 0.22 pm for MRRs 1-3, respectively. The curves in Figure \ref{fig:ThDM} show how $\Delta\lambda$ evolves with PUC distance $d_i$. For the distances present in the chip, the wavelength shift per driven phase predominantly falls within the range of $[0.2, 0.5]\frac{pm}{\pi}$. These findings align with the values derived from the TPM fitting. It is expected that when modelling with a single parameter, its value would tend towards the lower end of the range, considering the higher count of distant PUCs compared to those closer to the MRR within this hexagonal mesh.

\subsection{LR}

\begin{figure}[!t]
    \centering
    \includegraphics[width=\columnwidth]{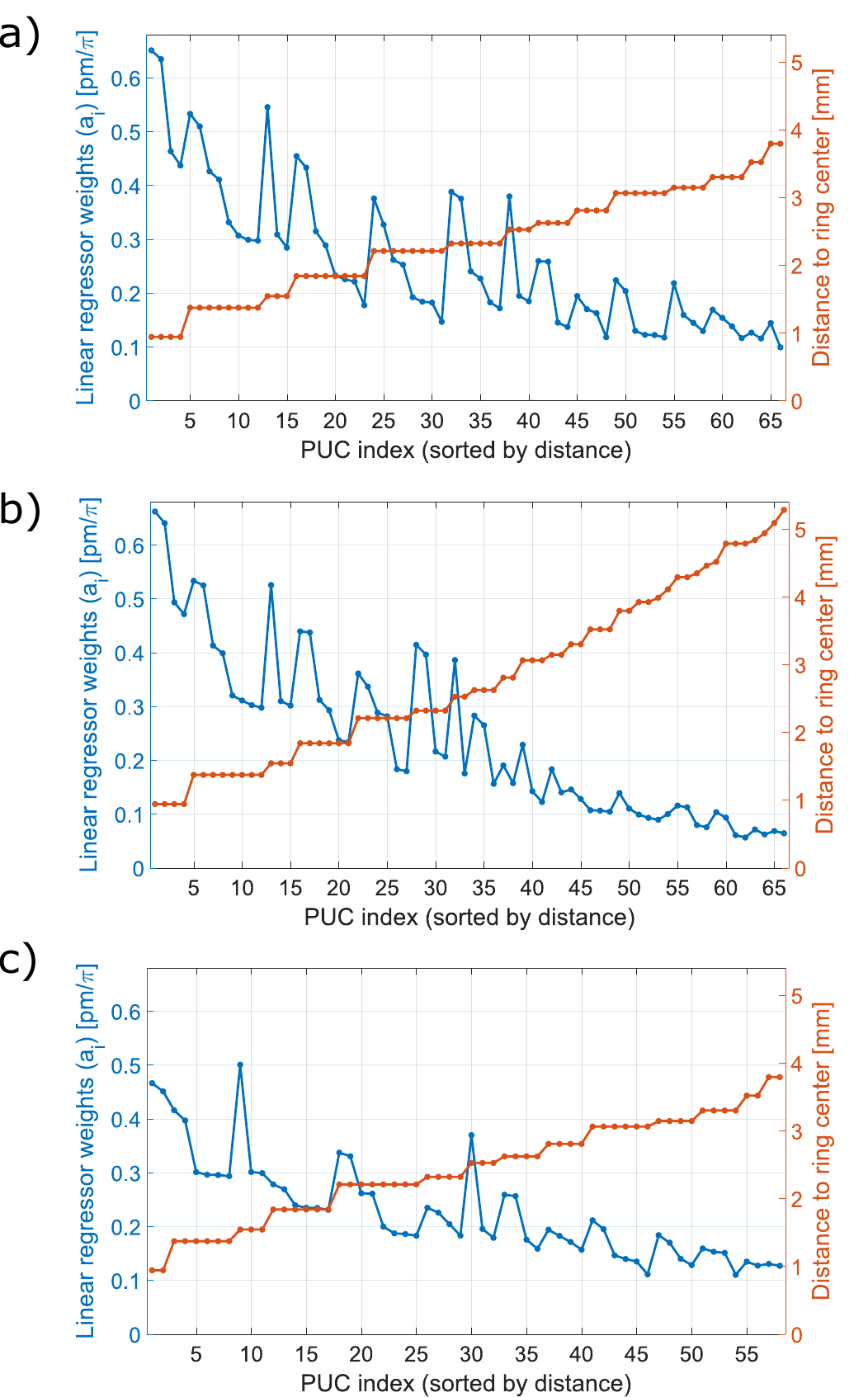}
    \caption{Ridge regression weights in $[pm/\pi]$ plotted alongside the distance to MRR center for each interfering PUC for a) MRR 1; b) MRR 2; and c) MRR 3.}
    \label{fig:LR}
\end{figure}

The LR weights were trained using ridge regression, with the regularization parameter optimized through five-fold cross-validation. The training RMSEs are determined to be 0.33 pm, 0.26 pm, and 0.20 pm, while testing RMSEs are measured at 0.34 pm, 0.27 pm, and 0.21 pm. The weights of the trained regression models are depicted in Figure \ref{fig:LR}. These weights are sorted based on their distance to the MRR and plotted in units of $[\frac{pm}{\pi}]$, for ease of comparison with the other models. Upon observing the plots, the figures indicate that the weights and distances are inversely correlated. This confirms that the black-box approach utilized during LR training leads to physically sensible outcomes - PUCs farther from the MRR tend to have a lower impact on the crosstalk-induced wavelength shift, and vice versa. 

The LR achieves the highest prediction accuracy compared to the TPM and the ThDM for all the MRRs. Additionally, a significant advantage of LR over ThDM is its ability to operate independently of precise chip layout knowledge. In other words, no information on the PUCs distance to the MRR center is needed for the model to accurately predict the crosstalk-induced $\Delta \lambda_{res}$. during the training, the LR model learns the contribution of each PUC to the accumulated crosstalk effect from experimental data only. As a result, it comprehensively grasps the behavior of each PUC, including fabrication variations and relative position (i.e., thermal distance) within the chip. This directly translates into notably accurate modeling.

As a final note, the regression weights predominantly fall within the range of $[0.1, 0.6] \frac{pm}{\pi}$, consistently aligning with ThDM across all the investigated MRRs.

\section{Training Set Size Analysis}\label{chap:training_size_analysis}
The performance of a model fitted on experimental data depends on the number of datapoints it is trained on. The testing performance typically improves with the number of training datapoints until it saturates at a level depending on the complexity of the model, which is strongly correlated with the number of free parameters optimized during the training. Models with more degrees of freedom tend to
require more datapoints until they can perform close to optimally and perform worse when only few datapoints are available.

\begin{figure}[!t]
    \centering
    \includegraphics[width=\columnwidth]{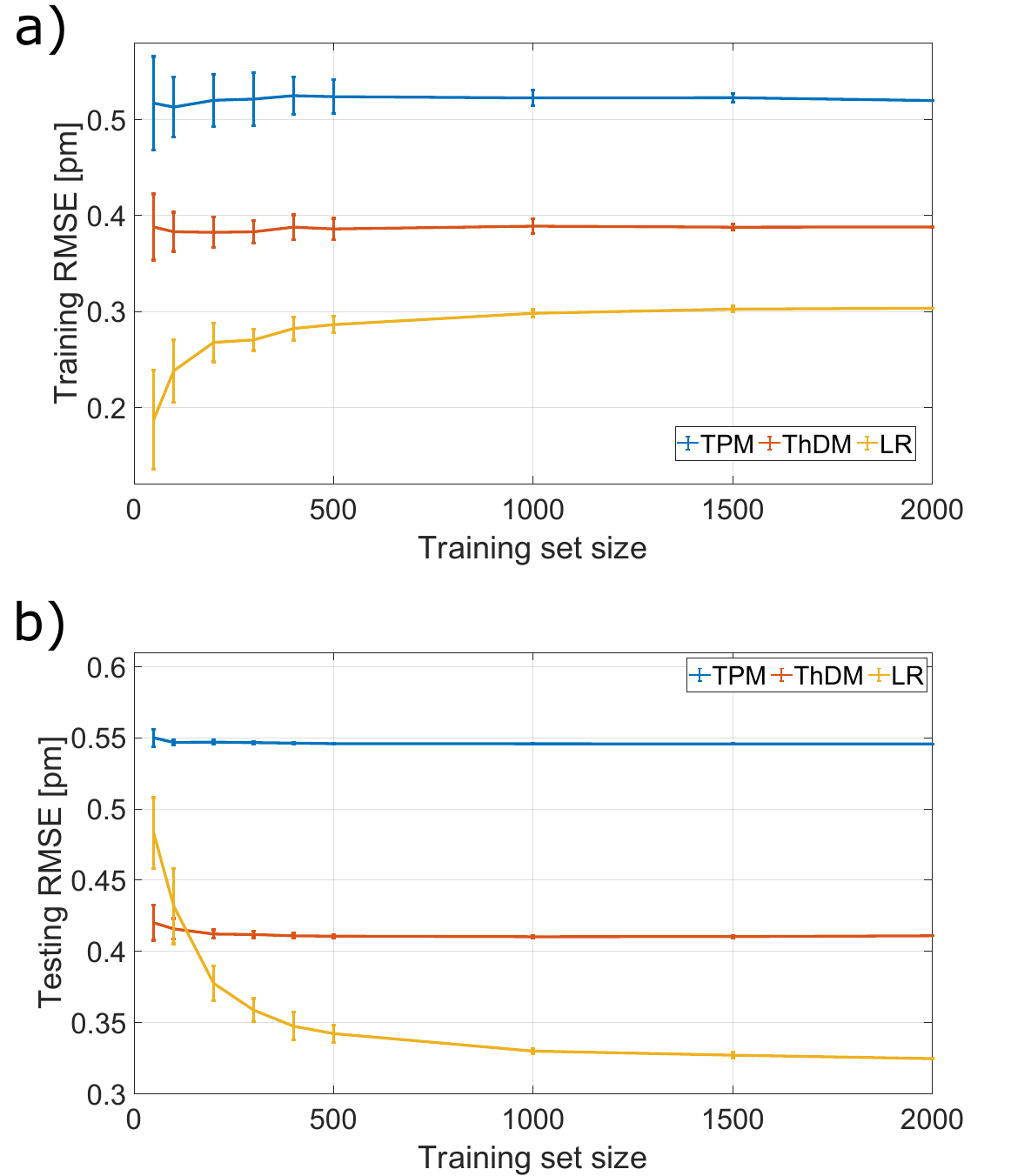}
    \caption{(a) Training and (b) Testing RMSEs for all the models for different training set sizes}
    \label{fig:Ltrain_size}
\end{figure}

The evolution of training and testing RMSEs as the size of the training set increases is shown in Figure \ref{fig:Ltrain_size}. For each training set size under investigation, 20 subsets of datapoints were selected randomly from the training set (with replacement) to perform the training. The error bars in the figures represent ±1 standard deviation from the mean for the 20 subsets. For all 3 models under investigation, both the training and testing performances remain within error bars after more than 2000 datapoints are used, indicating that using 4000 datapoints (80$\%$ of 5000) for training was more than enough for the models to reach their optimal performances.

Starting with the training RMSEs - for all 3 models, the error bars gradually get narrower as the training set size is increased. However, the mean training error does not change significantly for the TPM and the ThDM due to the fact that both models only fit very few parameters. On the other hand, the training error for LR gradually increases until the training set size is around 1000. This points towards overfitting for the linear regression model despite regularization when few experimental measurements are available for training, which is even clearer when the evolution of the testing RMSE is observed. For the LR, the testing error mirrors the training error, i.e. it starts off at a higher value at around 0.5 pm when 50 measurements are used and gradually decreases until it saturates at a level slightly higher than 0.3 pm. The fact that the training and testing errors mirror each other and converge to the same error level shows that with more experimental measurements for training, the generalization ability of the LR increases and its performance converges after more than 1000 measurements are used. As expected, the testing errors for the other models remain stable and slightly higher than the training errors. As a final point, the ThDM outperforms the data-driven LR model when only 100 or fewer experimental measurements are available, as the ThDM inherently uses information (PUC distances) that LR needs to learn from the measurements.

\section{Compensation performance}\label{chap:compensation_performance}

While LR shows lower modelling (both training and testing) errors, the physical perspective included in ThDM could offer better across-the-chip generalization performance. In other words, The ThDM trained for one of the MRRs can be straightforwardly translated to predict the crosstalk-induced $\Delta \lambda_{res}$ for any other MRR in the mesh, as it does not depend on the number of interfering PUCs and can extrapolate for distances not present in its training dataset. Therefore, the ThDM models are evaluated for predicting $\Delta \lambda_{res}$ for the MRRs they have not been trained on. The experimental evaluation is then reported for a small, randomly sampled subset of datapoints from each MRR's testing dataset.

\begin{figure}[!t]
    \centering
    \includegraphics[width=\columnwidth]{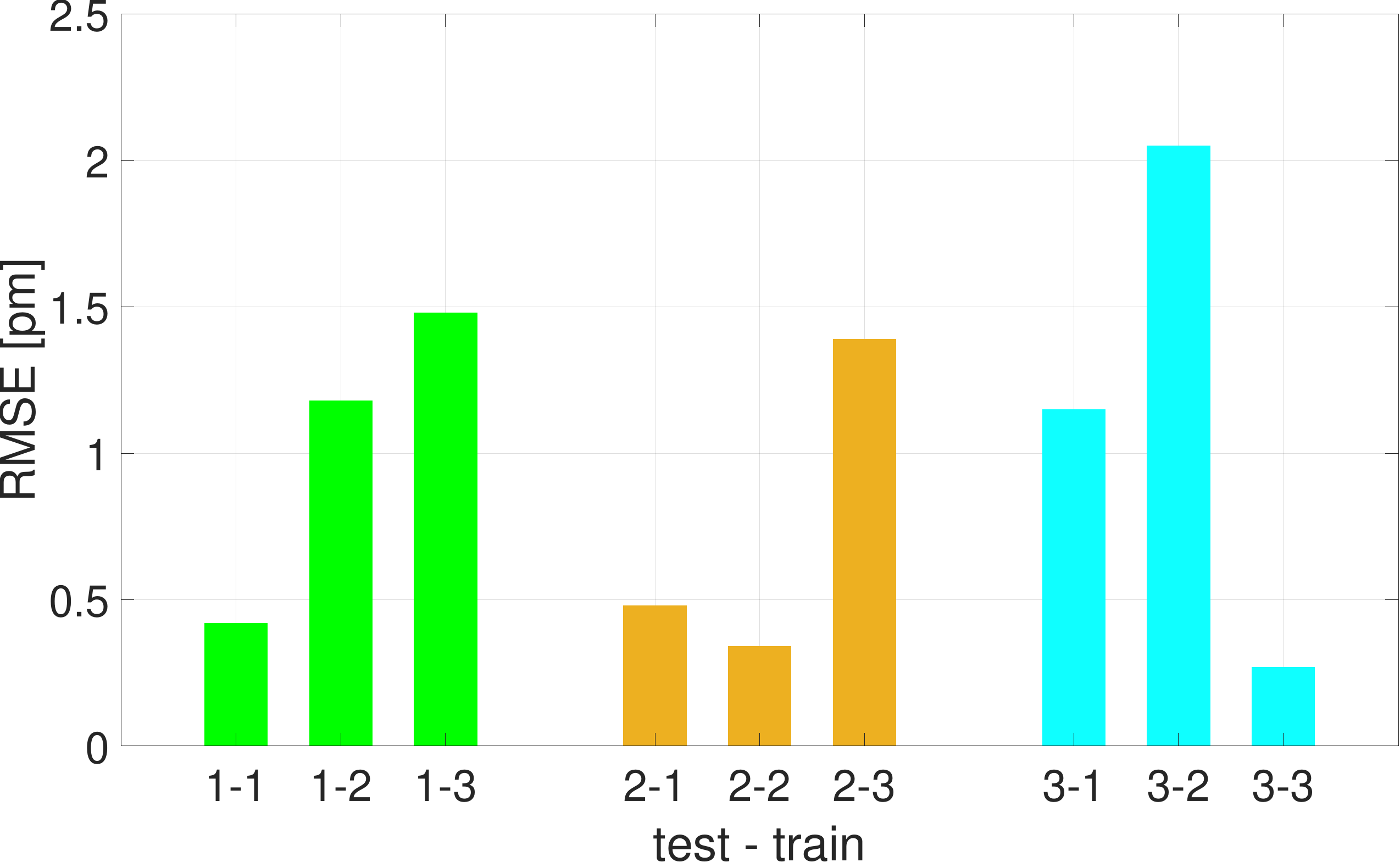}
    \caption{RMSE: $m-n$ refer to the ThDM trained for MRR $n$ evaluated for MRR $m$.}
    \label{fig:error_bars}
\end{figure}

RMSEs obtained when ThDM trained for MRR $n$ is evaluated on the testing dataset of MRR $m$ are given in Figure \ref{fig:error_bars}. Revealing RMSEs of 0.48 and 1.15 pm when employed to predict $\Delta \lambda_{res}$ for MRR 2 and MRR 3, respectively, the model trained for MRR 1 shows the best generalization performance.We relate this result to the position of MRR 1 in the mesh. According to the ThDM curves in Figure \ref{fig:ThDM}, the contribution to the effect of thermal crosstalk starts diminishing for PUCs located at distances beyond around $3$ mm from the MRR center for all 3 MRRs. Observing the distances present in the MRRs' datasets (red curves in Figure \ref{fig:LR}), MRR 1 has 54 out of 66 interfering PUCs located in the relevant distance range ($< 3$ mm), while MRR 2 has 13 PUCs less in that distance range. Therefore, MRR 1 is more likely to learn more accurately the contributions of the relevant interfering PUCs compared to MRR 2. MRR 3 is located symmetrically to MRR 1; however, its dataset includes 58 PUCs in total due to having the guiding PUCs phase shifts fixed throughout the measurements. Nonetheless, the guiding PUCs are located at distances in the range $[0.94, 2.32]$ mm from the center of MRR 3. Therefore, employing models for MRR 2 and MRR 3 to predict $\Delta \lambda_{res}$ for another MRRs can be expected to result in slightly higher RMSEs. 
Using the model for MRR 3 to predict $\Delta \lambda_{res}$ for MRR 1 and MRR 2 reveals RMSEs of 1.48 and 1.39 pm, respectively. Predicting $\Delta \lambda_{res}$ for MRR 1 and MRR 3 using the model for MRR 2 results in RMSEs of 1.18 and 2.05 pm, respectively.

For experimental compensation purposes, 6 random datapoints are sampled from each of the test datasets of all the MRRs under investigation. The wavelength shifts are then predicted by the ThDM models trained for each of the MRRs. Compensation for the crosstalk-induced $\Delta \lambda_{res}$ is demonstrated by appropriately adjusting the interfering PUCs' phase shifters directly on the MRRs to counteract for the model-predicted $\Delta \lambda_{res}$.

The additional phase shift that needs to be implemented to compensate for the predicted $\Delta \lambda_{res}$ is calculated by \eqref{eq:phi_comp}.

\begin{equation}
    \phi_{comp}= -2\pi \frac{\Delta \lambda_{res}}{FSR}
    \label{eq:phi_comp}
\end{equation}

Therefore, phase shifts of each of the 6 PUCs forming the MRR loop are modified by $\phi_{comp}' = \frac{1}{6} \phi_{comp}$ to counteract the effect of thermal crosstalk. The output power spectra are then measured and $\Delta \lambda_{res}$ after compensation is calculated as described in \ref{chap:setup&datasets}.

\begin{figure}
    \centering
    \includegraphics[width=\columnwidth]{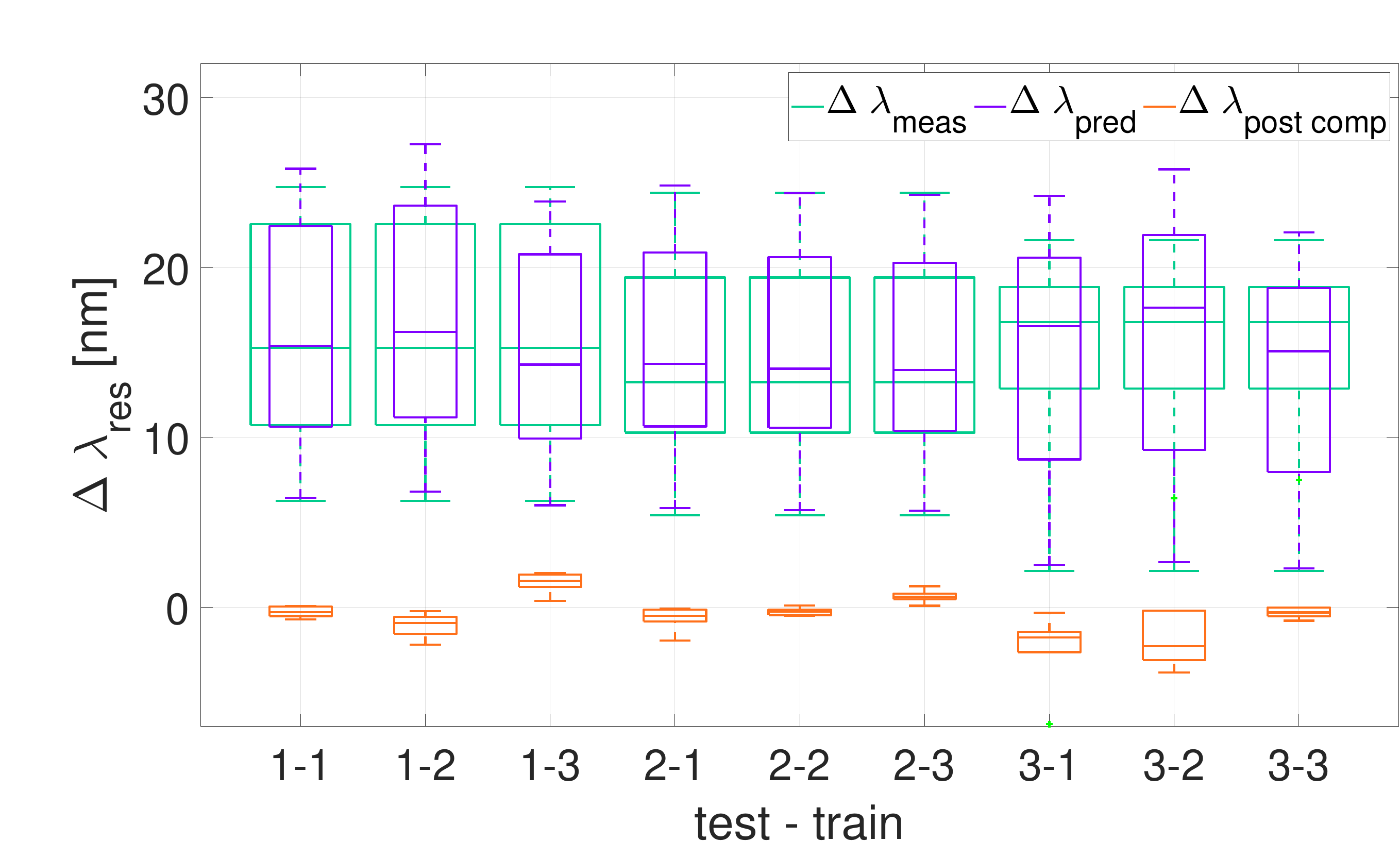}
    \caption{$\Delta \lambda_{res}$ boxplots showing performance of the ThDM when used to predict $\Delta \lambda_{res}$ for MRR m when trained on MRR n. The lines inside the box define the median, boxes correspond to the 25th and 75th percentiles, whiskers show the 10th and 90th percentiles for 6 testing samples. $\Delta \lambda_{meas}$ refers to measured $\Delta \lambda_{res}$, $\Delta \lambda_{pred}$ is $\Delta \lambda_{res}$ predicted by the ThDM, and $\Delta \lambda_{post \ comp}$ is $\Delta \lambda_{res}$ measured after the compensation.}
    \label{fig:boxplots}
\end{figure}

$\Delta \lambda_{res}$ before the compensation ($\Delta \lambda_{meas}$), $\Delta \lambda_{res}$ predicted by the models ($\Delta \lambda_{pred}$), and $\Delta \lambda_{res}$ after the compensation ($\Delta \lambda_{post \ comp}$) are given in boxplots in Figure \ref{fig:boxplots}.

Experimental validation confirms what is shown by the cross-predicting performance. The biggest discrepancy between the models trained for different MRRs is seen when the model for MRR 2 is used to predict $\Delta \lambda_{res}$ for MRR 3.

\section{Conclusion}\label{chap:conclusion}
We address the issue of thermal crosstalk which affects the programming accuracy of photonic programmable processors. Three distinct models are proposed and trained using experimental data. The models establish the correlation between the phases applied to the actuators on the photonic chip and the resulting resonance wavelength shift in the spectrum of the MRR programmed using the chip. Leveraging these modelling approaches, a model-based predictive crosstalk compensation technique is demonstrated by adjusting the phase shifters directly on the MRRs to counteract the crosstalk effect. This experimental procedure is replicated across three MRRs located on the same chip. LR model showed best modelling performance resulting in the lowest training and testing RMSEs, with the black-box training approach showing to yield reasonable physical outcomes. However, the training dataset size analysis showed that ThDM performs better with the limited training data. Furthermore, due to utilizing the mesh topology by incorporating the distances between interfering PUCs and MRRs, ThDM can be trained with one specific MRR and then applied to predict wavelength shifts for another MRR, paving the way towards more general off-line PIC modelling. This represents a valuable outcome of this research, since a general off-line method for programming PICs could lead to more accurate and less complex method to handle a highly present and performance-degrading impact of thermal crosstalk. Implementing crosstalk compensation could benefit scenarios requiring high phase sensitivity and contribute to the development of compact chip designs in the future.

\section*{Acknowledgments}
This work is supported by the Villum Young Investigator OPTIC-AI project (VIL29334), the Horizon Europe research and innovation project PROMETHEUS (grant n. 101070195), and ERC-StG No. 101076175 (LS-Photonics).

\bibliographystyle{IEEEtran}
\bibliography{JLT_PhotonicComputing_2024}

\vfill

\end{document}